\documentclass[10pt,english]{article}
\usepackage{times}
\usepackage[T1]{fontenc}
\usepackage[latin1]{inputenc}
\usepackage{amsmath}
\usepackage{amssymb}

\makeatletter

\providecommand{\tabularnewline}{\\}

\newcommand{\tensor}[1]{\stackrel{\leftrightarrow}{#1}}

\usepackage{slashed}

\usepackage{babel}
\makeatother
\begin{document}

\title{\textbf{Boson mass spectrum in $SU(4)_{L}\otimes U(1)_{Y}$ model
with exotic electric charges}}

\author{ADRIAN PALCU}

\date{\emph{Faculty of Exact Sciences - {}``Aurel Vlaicu'' University
Arad, Str. Elena Dr\u{a}goi 2, Arad - 310330, Romania}}

\maketitle
\begin{abstract}
The boson mass spectrum of the electro-weak \textbf{$SU(4)_{L}\otimes U(1)_{Y}$}
model with exotic electric charges is investigated by using the algebraical
approach supplied by the method of solving gauge models with high
symmetries. Our approach predicts for the boson sector a one-parameter
mass scale to be tuned in order to match the data obtained at LHC,
LEP, CDF. 


\end{abstract}

\section{Introduction}

In view of new experimental challenges - such as tiny massive neutrinos
and their oscillations or extra-neutral gauge bosons, to mention but
a few - the Standard Model (SM) \cite{key-1} - \cite{key-3} - based
on the gauge group $SU(3)_{C}\otimes SU(2)_{L}\otimes U(1)_{Y}$ that
undergoes in its electro-weak sector a spontaneous symmetry breakdown
(SSB) up to the electromagnetic universal one $U(1)_{em}$ - has to
be properly extended. 

One of the first and most popular such extensions - designed initially
to address another puzzle of the particle physics, namely the parity
non-conservation in weak interactions - was proposed in mid 70's by
Pati, Salam, Mohapatra and Senjanovic \cite{key-4} - \cite{key-7}
and is known as the (minimal) {}``left-right symmetric $SU(2)_{L}\otimes SU(2)_{R}\otimes U(1)_{B-L}$
model'' for electro-weak sector. Regarding the quark sector, which
is subject to the strong interaction, the {}``color'' gauge group
$SU(3)_{C}$ was replaced by $SU(4)_{C}$ . The so called Pati-Salam
model \cite{key-4} favors phenomenological consequences, such as
violation of the barion - lepton number in quark and proton decays
and the appearance of the exotic gauge mesons in semileptonic processes.
However, in the last decade, a new direction has emerged in the literature,
now being widely exploited. It resides in replacing the gauge group
$SU(2)_{L}$ in the electro-weak sector by the new $SU(4)_{L}$. Therefore,
the gauge symmetry $SU(3)_{C}\otimes SU(4)_{L}\otimes U(1)_{Y}$ (hereafter
3-4-1 model) was championed in a series of papers \cite{key-8} -
\cite{key-17} with notable results. According to the charge assignment,
different classes of 3-4-1 models were obtained by resorting to particular
methods \cite{key-18} - \cite{key-20} for discriminating among the
models.

It is worth noticing that both kinds of extensions above mentioned
seem to have common points, as it was recently emphasized in Refs.
\cite{key-21,key-22}. Plausible phenomenology can occur since, for
instance, the partial unification $SU(4)_{W}\otimes U(1)_{B-L}$ is
broken down to {}``minimal left-right'' symmetric model $SU(2)_{L}\otimes SU(2)_{R}\otimes U(1)_{B-L}$
either by four-dimensional Higgs mechanism or by inner or outer automorphisms
orbifolding in five dimensions \cite{key-22}. 

Here we deal with a particular class of 3-4-1 models, namely the one
including exotic electric charges \cite{key-8} - \cite{key-12} and
solve it by using the exact algebraical approach \cite{key-23} designed
for solving gauge models with high symmetries. Our main purpose is
to obtain for the boson sector a one-parameter mass scale to be tuned
in order to match the data obtained at LHC, LEP, CDF \cite{key-24}.
This goal is achieved by simply imposing a plausible physical criterion
on the resulting mass matrix. That is, in the low energy limit of
the model, the third neutral (diagonal) boson has to be decoupled
by the other two, since it comes from a higher breaking scale. 

The paper is organized as follows: Sec.2 briefly presents the main
steps of the general method of solving gauge models with high symmetries,
while Sec.3 deals with the 3-4-1 gauge model of interest which is
treated by using the above mentioned method. At the same time, some
estimates are given for different values of the free parameter $a$
in the resulting boson mass spectrum. Sec 4 sketches our conclusions.

\section{The General Method}

In this section we recall the main results of the general method \cite{key-23}
of treating $SU(n)_{L}\otimes U(1)_{Y}$ electro-weak gauge models,
in which a particular Higgs mechanism - based on a special parametrization
of the scalar sector - is embedded in order to properly break the
symmetry.

\subsection{$SU(n)_{L}\otimes U(1)_{Y}$ electro-weak gauge models}

In our general approach, the basic piece involved in the gauge symmetry
is the group $SU(n)$. Its two fundamental irreducible unitary representations
(irreps) $\mathbf{n}$ and $\mathbf{n^{*}}$ play a crucial role in
constructing different classes of tensors of ranks $(r,s)$ as direct
products like $(\otimes\mathbf{n})^{r}\otimes(\otimes\mathbf{n}^{*})^{s}$.
These tensors have $r$ lower and $s$ upper indices for which we
reserve the notation, $i,j,k,\cdots=1,\cdots,n$. As usually, we denote
the irrep $\rho$ of $SU(n)$ by indicating its dimension, ${\mathbf{n}}_{\rho}$.
The $su(n)$ algebra can be parameterized in different ways, but here
it is convenient to use the hybrid basis of Ref. \cite{key-23} consisting
of $n-1$ diagonal generators of the Cartan subalgebra, $D_{\hat{i}}$,
labeled by indices $\hat{i},\hat{j},...$ ranging from $1$ to $n-1$,
and the generators $E_{j}^{i}=H_{j}^{i}/\sqrt{2}$, $i\not=j$, related
to the off-diagonal real generators $H_{j}^{i}$ \cite{key-25,key-26}.
This way the elements $\xi=D_{\hat{i}}\xi^{\hat{i}}+E_{j}^{i}\xi_{i}^{j}\in su(n)$
are now parameterized by $n-1$ real parameters, $\xi^{\hat{i}}$,
and by $n(n-1)/2$ $c$-number ones, $\xi_{j}^{i}=(\xi_{i}^{j})^{*}$,
for $i\not=j$. The advantage of this choice is that the parameters
$\xi_{j}^{i}$ can be directly associated to the $c$-number gauge
fields due to the factor $1/\sqrt{2}$ which gives their correct normalization.
In addition, this basis exhibit good trace orthogonality properties,
\begin{equation}
Tr(D_{\hat{i}}D_{\hat{j}})=\frac{1}{2}\delta_{\hat{i}\hat{j}},\quad Tr(D_{\hat{i}}E_{j}^{i})=0\,,\quad Tr(E_{j}^{i}E_{l}^{k})=\frac{1}{2}\delta_{l}^{i}\delta_{j}^{k}\,.\label{Eq.1}\end{equation}
 When we consider different irreps, $\rho$ of the $su(n)$ algebra
we denote $\xi^{\rho}=\rho(\xi)$ for each $\xi\in su(n)$ such that
the corresponding basis-generators of the irrep $\rho$ are $D_{\hat{i}}^{\rho}=\rho(D_{\hat{i}})$
and $E_{j}^{\rho\, i}=\rho(E_{j}^{i})$.

The $U(1)_{Y}$ transformations are nothing else but phase factor
multiplications. Therefore - since the coupling constants $g$ for
$SU(n)_{L}$ and $g^{\prime}$ for the $U(1)_{Y}$ are assinged -
the transformation of the fermion tensor $L^{\rho}$ with respect
to the gauge group of the theory reads \begin{equation}
L^{\rho}\rightarrow U(\xi^{0},\xi)L^{\rho}=e^{-i(g\xi^{\rho}+g^{\prime}y_{ch}\xi^{0})}L^{\rho}\label{Eq.2}\end{equation}
 where $\xi=\in su(n)$ and $y_{ch}$ is the chiral hypercharge defining
the irrep of the $U(1)_{Y}$ group parametrized by $\xi^{0}$. For
simplicity, the general method deals with the character $y=y_{ch}g^{\prime}/g$
instead of the chiral hypercharge $y_{ch}$, but this mathematical
artifice does not affect in any way the results. Therefore, the irreps
of the whole gauge group $SU(n)_{L}\otimes U(1)_{Y}$ are uniquely
detemined by indicating the dimension of the $SU(n)$ tensor and its
character $y$ as $\rho=(\mathbf{n}_{\rho},y_{\rho})$.

In general, the spinor sector of our models has at least a part (usually
the leptonic one) which is put in pure left form using the charge
conjugation. Consequently this includes only left components, $L=\sum_{\rho}\oplus L^{\rho}$,
that transform according to an arbitrary reducible representation
of the gauge group. The Lagrangian density (Ld) of the free spinor
sector has the form \begin{equation}
{\mathcal{L}}_{S_{0}}=\frac{i}{2}\sum_{\rho}\overline{L^{\rho}}\tensor{\not\!\partial}L^{\rho}-\frac{1}{2}\sum_{\rho\rho'}\left(\overline{L^{\rho}}\chi^{\rho\rho^{\prime}}(L^{\rho^{\prime}})^{c}+h.c.\right).\label{Eq.3}\end{equation}
 Bearing in mind that each left-handed multiplet transforms as $L^{\rho}\rightarrow U^{\rho}(\xi^{0},\xi)L^{\rho}$
we understand that ${\mathcal{L}}_{S_{0}}$ remains invariant under
the global $SU(n)_{L}\otimes U(1)_{Y}$ transformations if the blocks
$\chi^{\rho\rho'}$ transform like $\chi^{\rho\rho^{\prime}}\rightarrow U^{\rho}(\xi^{0},\xi)\chi^{\rho\rho^{\prime}}(U^{\rho^{\prime}}(\xi^{0},\xi))^{T}$,
according to the representations $(\mathbf{n}_{\rho}\otimes\mathbf{n}_{\rho'},y_{\rho}+y_{\rho'})$
which generally are reducible. These blocks will give rise to the
Yukawa couplings of the fermions with the Higgs fields. The spinor
sector is coupled to the standard Yang-Mills sector constructed in
usual manner by gauging the $SU(n)_{L}\otimes U(1)_{Y}$ symmetry.
To this end we introduce the gauge fields $A_{\mu}^{0}=(A_{\mu}^{0})^{*}$
and $A_{\mu}=A_{\mu}^{+}\in su(n)$. Furthermore, the ordinary derivatives
are replaced in Eq. (\ref{Eq.3}) by the covariant ones, defined as
$D_{\mu}L^{\rho}=\partial_{\mu}L^{\rho}-ig(A_{\mu}^{\rho}+y_{\rho}A_{\mu}^{0})L^{\rho}$
thus arriving to the interaction terms of the spinor sector.

\subsection{Minimal Higgs Mechanism}

The scalar sector, organized as the so called minimal Higgs mechanism
(mHm) \cite{key-23}, is flexible enough to produce the SSB and, consequently,
generate masses for the plethora of particles and bosons in the model.
The scalar sector consists of $n$ Higgs multiplets $\phi^{(1)}$,
$\phi^{(2)}$, ... $\phi^{(n)}$ satisfying the orthogonality condition
$\phi^{(i)+}\phi^{(j)}=\phi^{2}\delta_{ij}$ in order to eliminate
the unwanted Goldstone bosons that could survive the SSB. $\phi$
is a gauge-invariant real scalar field while the Higgs multiplets
$\phi^{(i)}$ transform according to the irreps $(\mathbf{n},y^{(i)})$
whose characters $y^{(i)}$ are arbitrary numbers that can be organized
into the diagonal matrix \begin{equation}
Y=Diag\left(y^{(1)},y^{(2)},\cdots,y^{(n)}\right)\,.\label{Eq.4}\end{equation}
 The Higgs sector needs, in our approach, a parameter matrix \begin{equation}
\eta=Diag\left(\eta{}^{(1)},\eta{}^{(2)},...,\eta{}^{(n)}\right)\label{Eq.5}\end{equation}
 with the property ${\textrm{Tr}}(\eta^{2})=1-\eta_{0}^{2}$. It will
play the role of the metric in the kinetic part of the Higgs Ld which
reads \begin{equation}
\mathcal{L}_{H}=\frac{1}{2}\eta_{0}^{2}\partial_{\mu}\phi\partial^{\mu}\phi+\frac{1}{2}\sum_{i=1}^{n}\left(\eta{}^{(i)}\right)^{2}\left(D_{\mu}\phi^{(i)}\right)^{+}\left(D^{\mu}\phi^{(i)}\right)-V(\phi)\label{Eq.6}\end{equation}
 where $D_{\mu}\phi^{(i)}=\partial_{\mu}\phi^{(i)}-ig(A_{\mu}+y^{(i)}A_{\mu}^{0})\phi^{(i)}$
are the covariant derivatives of the model and $V(\phi)$ is the scalar
potential generating the SSB of the gauge symmetry \cite{key-23}.
This is assumed to have an absolute minimum for $\phi=\langle\phi\rangle\not=0$
that is, $\phi=\langle\phi\rangle+\sigma$ where $\sigma$ is the
unique surviving physical Higgs field. Therefore, one can always define
the unitary gauge where the Higgs multiplets, $\hat{\phi}^{(i)}$
have the components \begin{equation}
\hat{\phi}_{k}^{(i)}=\delta_{ik}\phi=\delta_{ik}(\langle\phi\rangle+\sigma)\,.\label{Eq.7}\end{equation}

This will be of great importance when the fermion masses will be computed,
due to the fact that the fermion mass terms - provided by Eq. (\ref{Eq.3})
via this mHm - exhibit the Yukawa traditional form only when the theory
is boosted towards the unitary gauge.

\subsection{Neutral bosons}

A crucial goal is now to find the physical neutral bosons with well-defined
properties. This must start with the separation of the electromagnetic
potential $A_{\mu}^{em}$ corresponding to the surviving $U(1)_{em}$
symmetry. We have shown that the one-dimensional subspace of the parameters
$\xi^{em}$ associated to this symmetry assumes a particular direction
in the parameter space $\lbrace\xi^{0},\xi^{\hat{i}}\rbrace$ of the
whole Cartan subalgebra. This is uniquely determined by the $n-1$
- dimensional unit vector $\nu$ and the angle $\theta$ giving the
subspace equations $\xi^{0}=\xi^{em}\cos\theta$ and $\xi^{\hat{i}}=\nu_{\hat{i}}\xi^{em}\sin\theta$.
On the other hand, since the Higgs multiplets in unitary gauge remain
invariant under $U(1)_{em}$ transformations, we must impose the obvious
condition $D_{\hat{i}}\xi^{\hat{i}}+Y\xi^{0}=0$ which yields \begin{equation}
Y=-D_{\hat{i}}\nu^{\hat{i}}\tan\theta\equiv-(D\cdot\nu)\tan\theta\,.\label{Eq.8}\end{equation}
 In other words, the new parameters $(\nu,\theta)$ determine all
the characters $y^{(i)}$ of the irreps of the Higgs multiplets. For
this reason these will be considered the principal parameters of the
model and therefore one deals with $\theta$ and $\nu$ (which has
$n-2$ independent components) instead of $n-1$ parameters $y^{(i)}$.

Under these circumstances, the generating mass term \begin{equation}
\frac{g^{2}}{2}\langle\phi\rangle^{2}Tr\left[\left(A_{\mu}+YA_{\mu}^{0}\right)\eta^{2}\left(A^{\mu}+YA^{0\mu}\right)\right]\,,\label{Eq.9}\end{equation}
 depends now on the parameters $\theta$ and $\nu_{\hat{i}}$. The
neutral bosons in Eq. \ref{Eq.9} being the electromagnetic field
$A_{\mu}^{em}$ and the $n-1$ new ones, $A_{\mu}^{'\hat{i}}$, which
are the diagonal bosons remaining after the separation of the electromagnetic
potential \cite{key-23}.

This term straightforwardly gives rise to the masses of the non-diagonal
gauge bosons \begin{equation}
M_{i}^{j}=\frac{1}{2}g\left\langle \phi\right\rangle \sqrt{\left[\left(\eta^{(i)}\right)^{2}+\left(\eta^{(j)}\right)^{2}\right]}\,,\label{Eq.10}\end{equation}
 while the masses of the neutral bosons $A_{\mu}^{'\hat{i}}$ have
to be calculated by diagonalizing the matrix \begin{equation}
(M^{2})_{\hat{i}\hat{j}}=\langle\phi\rangle^{2}Tr(B_{\hat{i}}B_{\hat{j}})\label{Eq.11}\end{equation}
 where \begin{equation}
B_{\hat{i}}=g\left(D_{\hat{i}}+\nu_{\hat{i}}(D\cdot\nu)\frac{1-\cos\theta}{\cos\theta}\right)\eta,\label{Eq.12}\end{equation}
 As it was expected, $A_{\mu}^{em}$ does not appear in the mass term
and, consequently, it remains massless. The other neutral gauge fields
${A'}_{\mu}^{\hat{i}}$ have the non-diagonal mass matrix (\ref{Eq.11}).
This can be brought in diagonal form with the help of a new $SO(n-1)$
transformation, $A_{\mu}^{'\hat{i}}=\omega_{\cdot\;\hat{j}}^{\hat{i}\;\cdot}Z_{\mu}^{\hat{j}}$
, which leads to the physical neutral bosons $Z_{\mu}^{\hat{i}}$
with well-defined masses. Performing this $SO(n-1)$ transformation
the physical neutral bosons are completely determined. The transformation
\begin{eqnarray}
A_{\mu}^{0} & = & A_{\mu}^{em}\cos\theta-\nu_{\hat{i}}\omega_{\cdot\;\hat{j}}^{\hat{i}\;\cdot}Z_{\mu}^{\hat{j}}\sin\theta,\nonumber \\
A_{\mu}^{\hat{k}} & = & \nu^{\hat{k}}A_{\mu}^{em}\sin\theta+\left(\delta_{\hat{i}}^{\hat{k}}-\nu^{\hat{k}}\nu_{\hat{i}}(1-\cos\theta)\right)\omega_{\cdot\;\hat{j}}^{\hat{i}\;\cdot}Z_{\mu}^{\hat{j}}.\label{Eq.13}\end{eqnarray}
 which switches from the original diagonal gauge fields, $(A_{\mu}^{0},A_{\mu}^{\hat{i}})$
to the physical ones, $(A_{\mu}^{em},Z_{\mu}^{\hat{i}})$ is called
the generalized Weinberg transformation (gWt).

\subsection{Electric and Neutral Charges}

The next step is to identify the charges of the particles with the
coupling coefficients of the currents with respect to the above determined
physical bosons. Thus, we find that the spinor multiplet $L^{\rho}$
(of the irrep $\rho$) has the following electric charge matrix \begin{equation}
Q^{\rho}=g\left[(D^{\rho}\cdot\nu)\sin\theta+y_{\rho}\cos\theta\right],\label{Eq.14}\end{equation}
 and the $n-1$ neutral charge matrices \begin{equation}
Q^{\rho}(Z^{\hat{i}})=g\left[D_{\hat{k}}^{\rho}-\nu_{\hat{k}}(D^{\rho}\cdot\nu)(1-\cos\theta)-y_{\rho}\nu_{\hat{k}}\sin\theta\right]\omega_{\cdot\;\hat{i}}^{\hat{k}\;\cdot}\label{Eq.15}\end{equation}
 corresponding to the $n-1$ neutral physical fields, $Z_{\mu}^{\hat{i}}$.
All the other gauge fields, namely the charged bosons $A_{j\mu}^{i}$,
have the same coupling, $g/\sqrt{2}$, to the fermion multiplets.

\section{Solving the $SU(4)_{L}\otimes U(1)_{Y}$ model}

The general method - constructed in Ref. \cite{key-23} and briefly
presented in the above section - is based on the following assumptions
in order to give viable results when it is applied to concrete models:

({\small I}) the spinor sector must be put (at least partially) in
pure left form using the charge conjugation (see for details Appendix
B in Ref. \cite{key-23})

({\small II}) the minimal Higgs mechanism - with arbitrary parameters
$(\eta_{0},\eta)$ satisfying the condition $Tr(\eta^{2})=1-\eta_{0}^{2}$
and giving rise to traditional Yukawa couplings in unitary gauge -
must be employed

({\small III}) the coupling constant, $g$, is the same with the first
one of the SM

({\small IV}) at least one $Z$-like boson should satisfy the mass
condition $m_{Z}=m_{W}/\cos\theta_{W}$ established in the SM and
experimentally confirmed.

Bearing in mind all these necessary ingredients, we proceed to solving
the particular 3-4-1 model \cite{key-8} - \cite{key-12} by imposing
from the very beginning the set of parameters we will work with.

\subsection{Fermion representations}

In what follows we denote the irreps of the electro-weak model under
consideration here by $\rho=(\mathbf{n}_{\rho},y_{ch}^{\rho})$ indicating
the genuine chiral hypercharge $y_{ch}$ instead of $y$. Therefore,
the multiplets of the 3-4-1 model of interest here will be denoted
by $(\mathbf{n}_{color},\mathbf{n}_{\rho},y_{ch}^{\rho})$. 

With this notation, after little algebra involving Eqs. (\ref{Eq.14})
- (\ref{Eq.15}) and the versor setting $\nu_{1}=1$, $\nu_{2}=0$,
$\nu_{3}=0$ - Case 1 in Ref. \cite{key-23}, one finds the following
irreps of the spinor sector:

\textbf{Lepton families}\begin{equation}
\begin{array}{ccccc}
f_{\alpha L}=\left(\begin{array}{c}
e_{\alpha}^{c}\\
e_{\alpha}\\
\nu_{\alpha}\\
N_{\alpha}\end{array}\right)_{L}\sim(\mathbf{1,4},0)\end{array}\label{Eq.18}\end{equation}

\textbf{Quark families}\begin{equation}
\begin{array}{ccc}
Q_{iL}=\left(\begin{array}{c}
J_{i}\\
u_{i}\\
d_{i}\\
D_{i}\end{array}\right)_{L}\sim(\mathbf{3,4^{*}},-1/3) &  & Q_{3L}=\left(\begin{array}{c}
J_{3}\\
-b\\
t\\
T\end{array}\right)_{L}\sim(\mathbf{3},\mathbf{4},+2/3)\end{array}\label{Eq.19}\end{equation}
\begin{equation}
\begin{array}{ccc}
(b_{L})^{c},(d_{iL})^{c},(D_{iL})^{c}\sim(\mathbf{3},\mathbf{1},+1/3) &  & (t_{L})^{c},(u_{iL})^{c},(T_{L})^{c}\sim(\mathbf{3},\mathbf{1},-2/3)\end{array}\label{Eq.20}\end{equation}
\begin{equation}
\begin{array}{ccccccccc}
(J_{3L})^{c}\sim(\mathbf{3,1},-5/3) &  &  &  &  &  &  &  & (J_{iL})^{c}\sim(\mathbf{3,1},+4/3)\end{array}\label{Eq.21}\end{equation}
 with $\alpha=1,2,3$ and $i=1,2$. With this, the conditions (I)
and (III) are fulfilled.

In addition, the connection between the $\theta$ angle of our method
and $\theta_{W}$(the Weinberg angle from SM) was inferred \cite{key-20}:
$\sin\theta=2\sin\theta_{W}$ along with the coupling relation: $g^{\prime}/g=\sin\theta_{W}/\sqrt{1-4\sin^{2}\theta_{W}}$. 

In the representations presented above we assumed, like in majority
of the papers in the literature, that the third generation of quarks
transforms differently from the other two ones. This could explain
the unusual heavy masses of the third generation of quarks, and especially
the uncommon properties of the top quark. The capital letters $J$
denote the exotic quarks included in each family. With this assignment
the fermion families cancel all the axial anomalies by just an interplay
between them, although each family remains anomalous by itself. Note
that one can add at any time sterile neutrinos - \emph{i.e.} right-handed
neutrinos $\nu_{\alpha R}\sim(\mathbf{1,1},0)$ - that could pair
in the neutrino sector of the Ld with left-handed ones in order to
eventually generate tiny Dirac or Majorana masses by means of an adequate
see-saw mechanism. These sterile neutrinos do not affect anyhow the
anomaly cancelation, since all their charges are zero. Moreover, their
number is not restricted by the number of flavors in the model.

\subsection{Boson mass spectrum}

Subsequently, we will use the standard generators $T_{a}$ of the
$su(4)$ algebra. Under these circumstances, the Hermitian diagonal
generators of the Cartan subalgebra are, in order, $D_{1}=T_{3}=\frac{1}{2}Diag(1,-1,0,0)$,
$D_{2}=T_{8}=\frac{1}{2\sqrt{3}}Diag(1,1,-2,0)$, and $D_{3}=T_{15}=\frac{1}{2\sqrt{6}}Diag(1,1,1,-3)$
respectively. In this basis, the gauge fields are $A_{\mu}^{0}$ and
$A_{\mu}\in su(4)$, that is \begin{equation}
A_{\mu}=\frac{1}{2}\left(\begin{array}{cccc}
D_{\mu}^{1} & \sqrt{2}X_{\mu} & \sqrt{2}X_{\mu}^{\prime} & \sqrt{2}K_{\mu}\\
\\\sqrt{2}X_{\mu}^{*} & D_{\mu}^{2} & \sqrt{2}W_{\mu} & \sqrt{2}K_{\mu}^{\prime}\\
\\\sqrt{2}X_{\mu}^{\prime*}{} & \sqrt{2}W_{\mu}^{*} & D_{\mu}^{3} & \sqrt{2}Y_{\mu}\\
\\\sqrt{2}K_{\mu}^{*} & \sqrt{2}K_{\mu}^{\prime*} & \sqrt{2}Y_{\mu}^{*} & D_{\mu}^{4}\end{array}\right),\label{Eq.22}\end{equation}
 with neutral diagonal bosons: $D_{\mu}^{1}=A_{\mu}^{3}+A_{\mu}^{8}/\sqrt{3}+A_{\mu}^{15}/\sqrt{6}$,
$D_{\mu}^{2}=-A_{\mu}^{3}+A_{\mu}^{8}/\sqrt{3}+A_{\mu}^{15}/\sqrt{6}$,
$D_{\mu}^{3}=-2A_{\mu}^{8}/\sqrt{3}+A_{\mu}^{15}/\sqrt{6}$, and $D_{\mu}^{4}=-3A_{\mu}^{15}/\sqrt{6}$
respectively. 

Apart from the charged Weinberg bosons, $W$, there are new charged
bosons, $K$, $K^{\prime}$, $X$, $X^{\prime}$ and $Y$. Note that
$X$ is doubly charged coupling different chiral states of the same
charged lepton (the so called ''bilepton''), while $Y$ is neutral.

The masses of both the neutral and charged bosons depend on the choice
of the matrix $\eta$ whose components are free parameters. Here it
is convenient to assume the following matrix \begin{equation}
\eta^{2}=(1-\eta_{0}^{2})Diag\left(1-c,\frac{1}{2}a-b,\frac{1}{2}a+b,c-a\right),\label{Eq.23}\end{equation}
 where, for the moment, $a$,$b$ and $c$ are arbitrary non-vanishing
real parameters. Obviously, $\eta_{0},c\in[0,1)$, $a\in(0,c)$ and
$b\in(-a,+a)$. Note that with this parameter choice the condition
(II) is accomplished. 

Under these circumstances, the mass spectrum of the off-diagonal bosons
according to (\ref{Eq.10}) are \begin{eqnarray}
{m}^{2}(W) & = & m^{2}a,\label{Eq.24}\\
{m}^{2}(X) & = & m^{2}\left(1-c+\frac{{\textstyle 1}}{2}a-b\right),\label{Eq.25}\\
{m}^{2}(X') & = & m^{2}\left(1-c+\frac{{\textstyle 1}}{2}a+b\right),\label{Eq.26}\\
{m}^{2}(K) & ={} & m^{2}(1-a),\label{Eq.27}\\
{m}^{2}(K') & = & m^{2}\left(c-\frac{{\textstyle 1}}{2}a-b\right),\label{Eq.28}\\
{m}^{2}(Y) & = & m^{2}\left(c-\frac{{\textstyle 1}}{2}a+b\right).\label{Eq.29}\end{eqnarray}
 while the mass matrix of the neutral bosons is given by Eq. (\ref{Eq.11})
\begin{equation}
M^{2}=m^{2}\left(\begin{array}{ccc}
\frac{{\textstyle 1}}{{\textstyle \cos^{2}\theta}}\left(1-c+\frac{{\textstyle 1}}{2}a-b\right) & \frac{{\textstyle 1}}{{\textstyle \sqrt{3}\cos\theta}}\left(1-c-\frac{1}{{\textstyle 2}}a+b\right) & \frac{{\textstyle 1}}{{\textstyle \sqrt{6}\cos\theta}}\left(1-c-\frac{{\textstyle 1}}{{\textstyle 2}}a+b\right)\\
\\\frac{{\textstyle 1}}{{\textstyle \sqrt{3}\cos\theta}}\left(1-c-\frac{{\textstyle 1}}{{\textstyle 2}}a+b\right) & \frac{{\textstyle 1}}{{\textstyle 3}}\left(1-c+\frac{{\textstyle 5}}{{\textstyle 2}}a+3b\right) & \frac{{\textstyle 1}}{{\textstyle 3\sqrt{2}}}\left(1-c-\frac{{\textstyle 1}}{{\textstyle 2}}a-3b\right)\\
\\\frac{{\textstyle 1}}{{\textstyle \sqrt{6}\cos\theta}}\left(1-c-\frac{{\textstyle 1}}{{\textstyle 2}}a+b\right) & \frac{{\textstyle 1}}{{\textstyle 3\sqrt{2}}}\left(1-c-\frac{{\textstyle 1}}{{\textstyle 2}}a-3b\right) & \frac{{\textstyle 1}}{{\textstyle 6}}\left(1+8c-8a\right)\end{array}\right)\label{Eq.30}\end{equation}
 with $m^{2}=g^{2}\left\langle \phi\right\rangle ^{2}(1-\eta_{0}^{2})/4$
throughout this paper. In order to fulfil the requirement (IV), the
above matrix has to admit $m^{2}a/{\textstyle \cos^{2}\theta}_{W}$
as eigenvalue. 

Now, one can enforce some other phenomenological assumptions. First
of all, it is natural to presume that the third neutral (diagonal
boson) $Z^{\prime\prime}$ should be considered much heavier than
its companions, so that it decouples form their mixing as the symmetry
is broken to $SU(3)$. For this purpose a higher breaking scale is
responsable, that at the same time supplies mass to $Z^{\prime\prime}$.

Therefore one can well consider $M_{13}^{2}=M_{23}^{2}=M_{31}^{2}=M_{32}^{2}$
in the matrix (\ref{Eq.30}). This gives rise to the condition

\begin{equation}
b=-\left(\frac{\sqrt{3}-\sqrt{1-4\sin^{2}\theta_{W}}}{\sqrt{3}+3\sqrt{1-4\sin^{2}\theta_{W}}}\right)\left(1-c-\frac{1}{2}a\right),\label{Eq.31}\end{equation}
which naturally leads to \begin{equation}
1-c\simeq\frac{{\textstyle 1}}{{\textstyle 2}}a\label{Eq.32}\end{equation}
in order to vanish the above terms. 

Hence, with the plausible physical condition assumed, Eq. (\ref{Eq.30})
looks like

\begin{equation}
M^{2}=m^{2}\left(\begin{array}{ccccc}
\frac{1}{{\textstyle \cos^{2}\theta}}(a-b) &  & \frac{1}{{\textstyle 2\sqrt{3}\cos\theta}}b &  & 0\\
\\\frac{1}{{\textstyle 2\sqrt{3}\cos\theta}}b &  & a+b &  & 0\\
\\0 &  & 0 &  & \frac{{\textstyle 1}}{{\textstyle 2}}(3-4a)\end{array}\right)\label{Eq.33}\end{equation}

Let us observe that the condition (IV) - $Det\left|M^{2}-m^{2}a/\cos^{2}\theta_{W}\right|=0$
- is fulfilled if and only if $b=\frac{3}{2}a\tan^{2}\theta_{W}$,
resulting from diagonalization of the remaning part of the matrix
(\ref{Eq.33}). Therefore, one finally remains with only one parameter
- say $a$. 

In addition, for there are terms which become singular for $\cos\theta=0$
which corresponds to the value $\sin^{2}\theta_{W}=1/4$ the Weinberg
angle is restricted in our model so that $\sin^{2}\theta_{W}$ less
than $1/4$, which is in good accord to experimental measurements
on it \cite{key-24}.

Obviously, $Z$ is the neutral boson of the SM, while $Z^{\prime}$
is a new neutral boson of this model (also occuring in 3-3-1 models)
whose mass comes form $Tr(M^{2})={m}^{2}(Z)+{m}^{2}(Z^{\prime})+{m}^{2}(Z^{\prime\prime})$. 

\begin{eqnarray}
{m}^{2}(W) & = & m^{2}a,\label{Eq.34}\\
{m}^{2}(X) & = & m^{2}a\left(1-\frac{3}{2}\tan^{2}\theta_{W}\right),\label{Eq.35}\\
{m}^{2}(X') & = & m^{2}a\left(1+\frac{3}{2}\tan^{2}\theta_{W}\right),\label{Eq.36}\\
{m}^{2}(K) & ={} & m^{2}(1-a),\label{Eq.37}\\
{m}^{2}(K') & = & m^{2}\left[1-a\left(1+\frac{3}{2}\tan^{2}\theta_{W}\right)\right],\label{Eq.38}\\
{m}^{2}(Y) & = & m^{2}\left[1-a\left(1-\frac{3}{2}\tan^{2}\theta_{W}\right)\right],\label{Eq.39}\\
{m}^{2}(Z) & = & m^{2}a/\cos^{2}\theta_{W},\label{Eq.40}\\
{m}^{2}(Z^{\prime}) & = & {m}^{2}a\left[1+\frac{3\sin^{2}\theta_{W}\left(1-2\sin^{2}\theta_{W}\right)}{1-4\sin^{2}\theta_{W}}\right],\label{Eq.41}\\
{m}^{2}(Z^{\prime\prime}) & = & m^{2}\frac{1}{2}(3-4a).\label{Eq.42}\end{eqnarray}

The mass scale is now just a matter of tuning the parameter $a$ in
accordance with the possible values for $\left\langle \phi\right\rangle $.
Obviously, parameter $a$ has now to be upper limited, so that $a\in(0,0.7)$
in order to have only positive values for the above experessions of
the squared boson masses.

\subsection{Numerical estimates}

In order to allow for a high breaking scale in the model $\langle\phi\rangle\geq1$TeV
and keep at the same time consistency with low energy phenomenology
of the SM our solution favors the case with $a\rightarrow0$ and $c\rightarrow1$.
However, assuming that $m(W)\simeq84.4$GeV and $m(Z)\simeq91.2$GeV
and $\sin^{2}\theta_{W}\simeq0.223$ (values taken from PDG 2008 -
Ref.\cite{key-24}), our approach predicts the exact masses at tree
level for the following bosons (according to Eqs. (\ref{Eq.35}),
(\ref{Eq.36}) and (\ref{Eq.41})): $m(X)\simeq63.7$GeV, $m(X^{\prime})\simeq100.9$GeV
and $m(Z^{\prime})\simeq177.6$GeV. Note that all these results are
independent of the unique mass scale of the model ($m$ in our notation). 

\begin{table}

\caption{Boson masses }

\begin{tabular}{ccccccccccc}
\hline 
$a$ \textbackslash{} Mass (GeV) &
&
$m=\frac{m(W)}{\sqrt{a}}$&
&
$m(Z^{\prime\prime})$&
&
$m(Y)$&
&
$m(K)$&
&
$m(K^{\prime})$\tabularnewline
\hline
&
&
&
&
&
&
&
&
&
&
\tabularnewline
$a=0.5$&
&
$119.36$&
&
$59.68$&
&
$100.9$&
&
$84.4$&
&
$63.7$\tabularnewline
&
&
&
&
&
&
&
&
&
&
\tabularnewline
$a=0.2$&
&
$188.72$&
&
$207.6$&
&
$177.6$&
&
$168.8$&
&
$159.5$\tabularnewline
&
&
&
&
&
&
&
&
&
&
\tabularnewline
$a=0.05$&
&
$377.45$&
&
$528.4$&
&
$372.0$&
&
$367.9$&
&
$363.7$\tabularnewline
&
&
&
&
&
&
&
&
&
&
\tabularnewline
$a=0.02$&
&
$596.80$&
&
$871.3$&
&
$593.4$&
&
$590.8$&
&
$588.2$\tabularnewline
&
&
&
&
&
&
&
&
&
&
\tabularnewline
$a=0.007$&
&
$1008.7$&
&
$1499.0$&
&
$1006.7$&
&
$1005.1$&
&
$1003.6$\tabularnewline
&
&
&
&
&
&
&
&
&
&
\tabularnewline
$a=0.005$&
&
$1193.6$&
&
$1778.5$&
&
$1191.9$&
&
$1190.6$&
&
$1189.3$\tabularnewline
&
&
&
&
&
&
&
&
&
&
\tabularnewline
$a=0.002$&
&
$1887.2$&
&
$2823.3$&
&
$1886.1$&
&
$1885.3$&
&
$1884.5$\tabularnewline
&
&
&
&
&
&
&
&
&
&
\tabularnewline
$a=0.0007$&
&
$3190.0$&
&
$4780.5$&
&
$3189.3$&
&
$3188.8$&
&
$3188.4$\tabularnewline
&
&
&
&
&
&
&
&
&
&
\tabularnewline
$a=0.0005$&
&
$3774.5$&
&
$5658.0$&
&
$3773.9$&
&
$3773.5$&
&
$3773.2$\tabularnewline
&
&
&
&
&
&
&
&
&
&
\tabularnewline
$a\rightarrow0$&
&
$m$&
&
$1.2m$&
&
$m$&
&
$m$&
&
$m$\tabularnewline
&
&
&
&
&
&
&
&
&
&
\tabularnewline
\hline 
&
&
&
&
&
&
&
&
&
&
\tabularnewline
\end{tabular}
\end{table}

The more heavier bosons are - roughly speaking - in the following
hierarchy mass $m(Z^{\prime\prime})>m(Y)>m(K)>m(K^{\prime})$ for
small values of the parameter $a$. If one desires a more detailed
estimate for the latter bosons, depending on the possible values of
the free parameter (and consequently on the mass scale), we present
a plot of their masses dependence on $a$ for a given $m$. That is,
we represented the functions $m(Z^{\prime\prime})\sim\sqrt{1.5-2a}$,
$m(Y)\sim\sqrt{1-0.57a}$, $m(K)\sim\sqrt{1-a}$ and $m(K^{\prime})\sim\sqrt{1-1.43a}$
(see Figure). 

If the mass scale of the model ($m$ in our notation), lies in the
TeV region - which is the current energy involved in LHC data - or
even in a higher one, the three latter masses become almost degenerate.
More precisely, if $m\simeq1$TeV, from Eq.(\ref{Eq.34}) $a\simeq0.00712336$
is inferred. Hence, $m(Z^{\prime\prime})\simeq1.485$TeV and $m(Y)\simeq0.998$TeV,
$m(K)\simeq0.996$TeV and $M(K^{\prime})\simeq0.995$TeV. Furthermore,
if the mass scale reaches GUT energies (specific to some see-saw mechanism)
these bosons - as expected - become degenerate.

We refer the reader to the following Table for some numerical results
in the boson sector of the given 3-4-1 model. A more accurate estimate
for the masses of these bosons and the relations among them (by a
more apropriate tuning of parameter $a$) will come, once the experimental
evidence of their phenomenology will be definitely bring to light
at LHC, LEP, CDF and other high energy accelerators in a near future.

\section{Conclusions}

In this work the boson mass spectrum of a 3-4-1 model with exotic
electric charges has been worked out within the framework of the general
method for solving gauge models with high symmetries \cite{key-23}
and analyzed by just tuning a unique free parameter $a$. In our approach
the latter is close related to the mass scale (and thus on the overall
breaking scale $\left\langle \phi\right\rangle $) in the manner $m\simeq84.4/\sqrt{a}$(GeV).
Plausible phenomenological predictions were made at a reasonable scale
arround $1$TeV to be confronted by data supplied by LHC, LEP, CDF
and other colliders. The method used is flexible enough in order to
allow even a different Higgs mechanism which could consist in re-defining
the scalar multiplets in the manner $\phi^{(i)}\rightarrow\eta{}^{(1)}\phi^{(i)}$
in order to match the traditional approach, with each new $\phi^{(i)}$
responsable for one particular step in breaking the symmetry. One
can further investigate this model by calculating the neutral currents
and specific decays, the oblique corrections $S,T,U$ \cite{key-21},
or conceive a suitable see-saw mechanism in order to generate tiny
masses for neutrinos. Breaking down the symmetry to the left-right
Pati-Salam purposal \cite{key-22} is also a worthy way for further
theoretical investigations. All these, however, are beyond the scope
of this paper and will be treated in a future work.

\end{document}